%% file: main.tex
\documentclass[10pt, conference]{IEEEtran}
\IEEEoverridecommandlockouts
\usepackage{todonotes}
\usepackage{tikz}
\usepackage{hyperref}
\usetikzlibrary{quantikz2}
\usepackage{float}
\usepackage{url}
\usepackage[thinlines]{easytable}
\usepackage{comment}
\usepackage[n,
 advantage,
 operators,
 sets,
 adversary,
 landau,
 probability,
 notions,
 logic,
 ff,
 mm,
 primitives,
 events,
 complexity,
 oracles,
 asymptotics,
 keys] {cryptocode}
\usepackage{tikzit}
\input{styles.tikzstyles}

\usepackage{cite}
\usepackage{amsmath,amssymb,amsfonts}
\usepackage{algorithmic}
\usepackage{graphicx}
\usepackage{textcomp}
\usepackage{booktabs}
\usepackage{xcolor}
\usepackage{mathtools}
\newcommand{\sqctrl}[1]{%
    \gate[style={fill=white, draw=black, inner sep=-1pt, minimum size=-1pt}]{}\vqw{#1}%
}

\begin{document}

\title{Architecture-aware Unitary Synthesis\\
\thanks{This work was funded by the Business Finland Quantum Computing
Campaign: The Enhanced Middleware for Quantum Systems (EM4QS) project.}
}

\author{\IEEEauthorblockN{1\textsuperscript{st} Frans Perkkola}
\IEEEauthorblockA{\textit{Dept. of Computer Science} \\
\textit{University of Helsinki}\\
Helsinki, Finland \\
frans.perkkola@helsinki.fi}
\and
\IEEEauthorblockN{2\textsuperscript{nd} Arianne Meijer-van de Griend}
\IEEEauthorblockA{\textit{Dept. of Computer Science} \\
\textit{University of Helsinki}\\
Helsinki, Finland \\
arianne.vandegriend@helsinki.fi}
\and
\IEEEauthorblockN{3\textsuperscript{rd} Jukka K. Nurminen}
\IEEEauthorblockA{\textit{Dept. of Computer Science} \\
\textit{University of Helsinki}\\
Helsinki, Finland \\
jukka.k.nurminen@helsinki.fi}
}

\maketitle

\begin{abstract}
We present a novel architecture-aware transpilation method for exact general unitary gate synthesis on superconducting quantum hardware. Our approach is tightly integrated with the optimized block-ZXZ decomposition, exploiting its recursive structure to make hardware-aware decisions at each level of the recursion rather than treating transpilation as an independent post-processing step. The method introduces three key techniques: a greedy qubit mapping strategy that minimizes pairwise distances between physical qubits, an adaptive Gray code selection combined with qubit swapping that optimizes the construction of uniformly controlled Rz gates for the target topology, and a heuristic for reducing CNOT gates by exploiting the structure of long-range CNOT ladders. We benchmark our method against TKET, Qiskit, and Pennylane on the 20-qubit IQM Garnet (square lattice) and the 156-qubit IBM Marrakesh (heavy-hex) architectures with qubit counts ranging from 3 to 11. Our method achieves CNOT count reductions of up to 36 percent on the IQM Garnet and up to 34 percent on the IBM Marrakesh compared to the best competing transpiler, while simultaneously achieving transpilation speedups of up to 553x. Furthermore, our method is the only one capable of transpiling circuits beyond 10 qubits within a 30-minute time limit across both architectures.

\end{abstract}


\section{Introduction}
Quantum computing promises significant speedups for a wide range of computational problems, from simulating quantum systems~\cite{lloyd1996universal, childs2018toward} to solving linear systems of equations~\cite{harrow2009quantum} and optimization problems~\cite{farhi2014quantum}. At the core of many quantum algorithms lies the implementation of generic unitary transformations, which must be compiled into sequences of elementary gates executable on physical quantum hardware. This compilation process, known as unitary synthesis, is a fundamental task in quantum computing, and its efficiency directly impacts the practical viability of quantum algorithms on near-term and early fault-tolerant quantum devices.
Generic unitary circuits appear as essential building blocks across numerous quantum algorithms and subroutines. In Hamiltonian simulation, the core task involves implementing the time-evolution operator $e^{-iHt}$, which is a unitary transformation that may require exact synthesis when product formula approximations are insufficient or when high precision is needed~\cite{lloyd1996universal, childs2018toward}. Some methods for quantum state preparation \cite{plesch-state-prep}, the process of initializing a quantum register into a desired quantum state, relies on the synthesizing unitary matrices into quantum circuits, and is a crucial subroutine in algorithms such as quantum machine learning~\cite{schuld2015introduction}, amplitude estimation~\cite{Brassard_2002}, and variational quantum eigensolvers~\cite{peruzzo2014variational}. Quantum signal processing (QSP) and its generalization, the quantum singular value transformation (QSVT)~\cite{gilyen2019quantum}, unify a broad class of quantum algorithms by applying polynomial transformations to block-encoded matrices, often requiring the implementation of carefully constructed unitary operators. Furthermore, block encoding techniques used in quantum linear algebra~\cite{low2019hamiltonian} necessitate the synthesis of structured unitary circuits. Other applications include the implementation of oracles in Grover's search algorithm~\cite{grover1996fast}, quantum walks~\cite{childs2009universal}, quantum error correction encoding circuits~\cite{gottesman1997stabilizer} and quantum phase estimation ~\cite{phase-estimation}, all of which may involve the compilation of nontrivial unitary transformations.

The theoretical lower bound for the number of CNOT gates required to implement a general $n$-qubit unitary is $ \frac{1}{4}(4^n - 3n - 1)
$~\cite{two-qubit-decomp}, and the current best known theoretical upper bound for exact synthesis is $\frac{22}{48} \cdot 4^n - \frac{3}{2} \cdot 2^n + \frac{5}{3}$
~\cite{blockzxz}, which is achieved by the optimized block-ZXZ decomposition. While these results characterize the gate complexity on a fully connected (all-to-all) qubit architecture, real quantum hardware imposes significant connectivity constraints. Superconducting quantum processors, such as the IQM Garnet~\cite{iqm-garnet} with its square lattice topology and the IBM Marrakesh with its heavy-hex topology, only allow two-qubit gates between physically adjacent qubits. Consequently, synthesized quantum circuits must be transpiled to respect these connectivity constraints, a process that typically introduces additional two-qubit gates and increases the overall circuit depth.

Existing quantum circuit transpilers, such as those provided by Qiskit~\cite{qiskit2023}, TKET~\cite{sivarajah2021tket}, and Pennylane~\cite{bergholm2018pennylane}, offer general-purpose transpilation routines that route arbitrary quantum circuits to a target architecture. However, these transpilers treat the transpilation as a post-processing step that is agnostic to the structure of the original synthesis. This separation between synthesis and transpilation leads to suboptimal results, as the transpiler cannot exploit the specific structure of the decomposition to minimize the overhead introduced by the routing process. Other works have explored topology-aware synthesis~\cite{topology, circuit-partitioning}, but either do not support exact synthesis, rely on slow numerical approximation methods~\cite{topology, bqskit}, or do not have publicly available implementations~\cite{circuit-partitioning, peephole}.
In this work, we address this gap by proposing an architecture-aware transpilation method that is tightly integrated with the optimized block-ZXZ unitary synthesis~\cite{blockzxz}. Our method exploits the recursive structure of the decomposition to make hardware-aware decisions at each level of the recursion, rather than treating transpilation as an independent post-processing step. The key insight is that the uniformly controlled $R_z$ $(UC)$ gates produced by the decomposition offer significant freedom in their implementation---specifically, the choice of Gray code and the physical placement of qubits---which can be optimized for the target hardware topology. We benchmark our method against TKET, Qiskit, and Pennylane on two representative hardware architectures: the 20-qubit IQM Garnet (square lattice) and the 156-qubit IBM Marrakesh (heavy-hex). Benchmarking against the BQSKit transpiler \cite{bqskit} was considered as well, but was not included in the end. Our results demonstrate significant reductions in both CNOT count and transpilation runtime across all tested qubit counts.

The key contributions of this article are as follows:
\begin{itemize}
    \item A novel architecture-aware transpilation method for exact general unitary synthesis, tightly integrated with the optimized block-ZXZ decomposition, that supports arbitrary quantum hardware topologies.
    \item Comprehensive benchmarking against TKET, Qiskit, and Pennylane on the IQM Garnet and IBM Marrakesh architectures, demonstrating CNOT count reductions of up to 36\% and transpilation speedups of up to 553x compared to existing methods.
\end{itemize}

\section{Background}
\label{sec:background}
This section briefly describes the optimized unitary synthesis method proposed in \cite{blockzxz} which we transpile for different superconducting quantum computer architectures. The method is based on the block-ZXZ method \cite{blockzxz_orig}, and uses the same optimizations used in the quantum Shannon decomposition (QSD) \cite{qsd}. We will briefly explain the block-ZXZ decomposition and the optimization methods employed for an all-to-all connected architecture.

\subsection{Block-ZXZ decomposition and demultiplexing}
The block-ZXZ decomposition decomposes a general unitary gate $U \in U(2^n)$ into the following structure:

\begin{align}
    U = \frac{1}{2}\begin{bmatrix}
        A_1 & 0 \\
        0 & A_2
    \end{bmatrix} (H \otimes I) \begin{bmatrix}
        I & 0 \\
        0 & B
    \end{bmatrix} (H \otimes I) \begin{bmatrix}
        I & 0 \\
        0 & C
    \end{bmatrix}
\end{align}
where $A_1, A_2, B$ and $C$ are some unitary matrices of dimension $2^{n-1} \times 2^{n-1}$ and $H$ is the Hadamard gate. \\

The resulting decomposition consists of unitary gates of the form $U = U_1 \oplus U_2$ called multiplexers. These multiplexers can be further decomposed, or demultiplexed, as follows:

\begin{align}
    \begin{bmatrix}
        U_1 & 0 \\
        0 & U_2
    \end{bmatrix} = \begin{bmatrix}
        V & 0 \\
        0 & V
    \end{bmatrix} \begin{bmatrix}
        D & 0 \\
        0 & D^\dagger{}
    \end{bmatrix} \begin{bmatrix}
        W & 0 \\
        0 & W
    \end{bmatrix}
\end{align}
where $D$ is a diagonal unitary gate. This technique can be used to demultiplex the gates $A, B$ and $C$, which gives the following circuit:

\[
\begin{quantikz}[row sep={0.6cm,between origins}, column sep = 0.1cm]
    & & \gate[2]{U} &\\
    & \qwbundle{} & &
\end{quantikz} = \begin{quantikz}[row sep={0.6cm,between origins}, column sep = 0.1cm]
   & & & \gate{R_z} & & \gate{H} & \gate{R_z}  & \gate{H} & & \gate{R_z} & &      \\
   & \qwbundle{} & \gate{W_C} & \sqctrl{-1} & \gate{V_C} & \gate{W_B} & \sqctrl{-1} & \gate{V_B} & \gate{W_A} & \sqctrl{-1} & \gate{V_A} &
\end{quantikz}
\]

It is clear that gate $V_C$ can be merged with $W_B$, and $V_B$ can be merged with $W_A$. Therefore, we have a decomposition from the general $n$-qubit unitary gate into four $(n-1)$-qubit unitary gates, with three uniformly controlled $R_z$ gates and two Hadamard gates. The uniformly controlled $R_z$ gates can be decomposed as in \cite{multiplexer-decomp}. The decomposition is applied recursively to each of the $(n-1)$-qubit unitary gates until only two-qubit unitaries are left, which will be decomposed with the methods proposed in \cite{two-qubit-decomp}. The details for the block-ZXZ decomposition and the demultiplexing can be found in \cite{blockzxz}.

\subsection{Optimization}
After applying the recursion until we have only two-qubit unitaries left, we decompose the unitaries using the techniques described in \cite{two-qubit-decomp}. The right-most two-qubit unitary can be decomposed up to a diagonal into the following circuit, which requires
only two CNOT gates:

\[
\begin{quantikz}[row sep={0.7cm,between origins}, column sep = 0.2cm]
    & \gate[2]{U} & \\
    & & 
\end{quantikz} = \begin{quantikz}[row sep={0.7cm,between origins}, column sep = 0.2cm]
    & \gate[2]{D} & \gate{U} & \ctrl{1} & \gate{R_x} & \ctrl{1} & \gate{U} & \\
    & & \gate{U} & \targ{} & \gate{R_z} & \targ{}& \gate{U} &
\end{quantikz}
\]
The diagonal gate can be then migrated through the controls of the uniformly controlled $R_z$ gates and merged with the next two-qubit unitary. This process is repeated until only one two-qubit unitary is left, which is then decomposed using three CNOT gates with the following circuit:

\[
\begin{quantikz}[row sep={0.7cm,between origins}, column sep = 0.2cm]
    & \gate[2]{U} & \\
    & & 
\end{quantikz} = \begin{quantikz}[row sep={0.7cm,between origins}, column sep = 0.2cm]
    & \gate{U} & \targ{} & \gate{R_z} & \ctrl{1} & & \targ{} & \gate{U} &\\
    &  \gate{U} & \ctrl{-1} & \gate{R_y} & \targ{} & \gate{R_y} & \ctrl{-1} & \gate{U} &
\end{quantikz}
\]
The single qubit unitary gates $U$ can be decomposed using the ZYZ-decomposition \cite{single-qubit-decomp}. \\

\begin{figure*}[!h]
    \centering
    $a)$\begin{quantikz}[row sep={0.6cm,between origins}, column sep = 0.1cm]
        & & \gate{R_z} & \gate{H} & \ctrl{1} & \gate{H} & \gate{R_z} & &\\
        & \gate[2]{W_C} & \sqctrl{-1} & \gate[2]{V_C} & \gate[2]{B} & \gate[2]{W_A} & \sqctrl{-1} & \gate[2]{V_A} & \\
        & & \sqctrl{-2} & & & & \sqctrl{-2} & &
    \end{quantikz} \\
    $b)$\begin{quantikz}[row sep={0.6cm,between origins}, column sep = 0.1cm]
        & \gate{R_z} & \targ{} & \gate{R_z} & \targ{} & \gate{R_z} & \targ{} & \gate{R_z} & \targ{} & \gate{H} & & \ctrl{1} & & \gate{H} & \targ{} & \gate{R_z} & \targ{} & \gate{R_z} & \targ{} & \gate{R_z} & \targ{} & \gate{R_z} &\\
        & \gate[2]{W_C} & & & \ctrl{-1} & & & & \ctrl{-1} & & \gate[2]{V_C} & \gate[2]{B} & \gate[2]{W_A} & & \ctrl{-1} & & & & \ctrl{-1} & & & \gate[2]{V_A} &\\
        & & \ctrl{-2} & & & & \ctrl{-2} & & & & & & & & & & \ctrl{-2} & & & & \ctrl{-2} & & 
    \end{quantikz} \\
    $c)$\begin{quantikz}[row sep={0.6cm,between origins}, column sep = 0.1cm]
        & \gate{R_z} & \targ{} & \gate{R_z} & \targ{} & \gate{R_z} & \targ{} & \gate{R_z} & \gate{H} & \control{} & & \ctrl{1} & & \control{} & \gate{H} & \gate{R_z} & \targ{} & \gate{R_z} & \targ{} & \gate{R_z} & \targ{} & \gate{R_z} &\\
        & \gate[2]{W_C} & & & \ctrl{-1} & & & &  & \ctrl{-1} & \gate[2]{V_C} & \gate[2]{B} & \gate[2]{W_A} & \ctrl{-1} &  & & & & \ctrl{-1} & & & \gate[2]{V_A} &\\
        & & \ctrl{-2} & & & & \ctrl{-2} & & & & & & & & & & \ctrl{-2} & & & & \ctrl{-2} & & 
    \end{quantikz} \\
    $d)$\begin{quantikz}[row sep={0.6cm,between origins}, column sep = 0.1cm]
        & \gate{R_z} & \targ{} & \gate{R_z} & \targ{} & \gate{R_z} & \targ{} & \gate{R_z} & \gate{H} & \ctrl{1} & \gate{H} & \gate{R_z} & \targ{} & \gate{R_z} & \targ{} & \gate{R_z} & \targ{} & \gate{R_z} &\\
        & \gate[2]{W_C} & & & \ctrl{-1} & & & &    & \gate[2]{\tilde{B}} &  & & & & \ctrl{-1} & & & \gate[2]{V_A} &\\
        & & \ctrl{-2} & & & & \ctrl{-2} & & & & & & \ctrl{-2} & & & & \ctrl{-2} & & 
    \end{quantikz}
    \caption{A three-qubit unitary example of the CNOT merging optimization procedure. a) Decompose only the $A$ and $C$ multiplexers. b) Decompose the uniformly controlled $R_z$ gates. c) Migrate the Hadamard gates through the CNOT targets. d) Merge $V_C, W_A$ and the CZ gates into the central gate $B$.}
    \label{fig:cnot-merging}
\end{figure*}
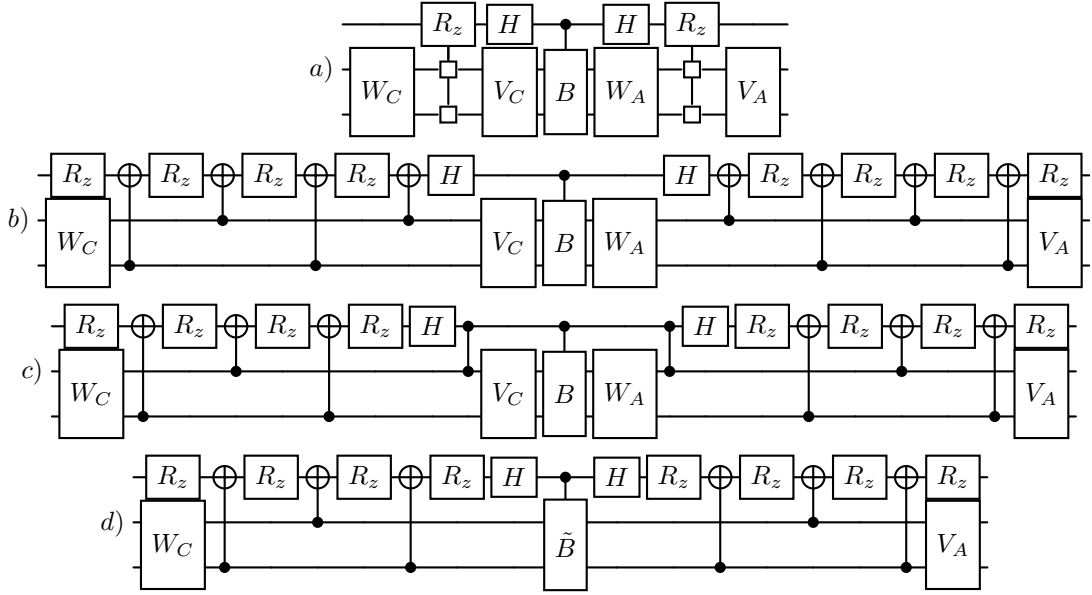

After the block-ZXZ decomposition, we first decompose only the left and right multiplexers ($A$ and $C$). We then have a circuit with two uniformly controlled $R_z$ gates and Hadamard gates, and the controlled gate $B$ in the middle. The two uniformly controlled $R_z$ gates can be decomposed such that both of the central Hadamard gates are next to a CNOT gate from a decomposed uniformly controlled $R_z$ gate. We can then move the Hadamard gates through the targets of the CNOT gates, turning them into CZ gates. The two CZ gates can be merged into the middle controlled gate $B$ together with the two $(n-1)$-qubit gates $V_C$ and $W_A$. The new central gate $\tilde{B}$ can be calculated as:

\begin{align}
    \tilde{B} &= \begin{bmatrix}
        W_AV_C & 0 \\
        0 & (Z \otimes I)W_ABV_C(Z\otimes I)
    \end{bmatrix}\\
    &= (CZ \otimes I) \begin{bmatrix}
        W_A & 0 \\
        0 & W_A
    \end{bmatrix} \begin{bmatrix}
        I & 0 \\
        0 & B
    \end{bmatrix} \begin{bmatrix}
        V_C & 0 \\
        0 & V_C
    \end{bmatrix} (CZ \otimes I)
\end{align}
and decomposed as a regular multiplexer. This saves two CNOT gates for every step of the recursion. Using these techniques we get the following CNOT count for the $n$-qubit unitary gate decomposition:
\begin{align}
    \label{block-zxz-bound}
    c_n \leq \frac{22}{48}\cdot 4^n-\frac{3}{2}\cdot2^n+\frac{5}{3}
\end{align}
which is the current best known upper bound for exact general unitary gate synthesis. An example of merging the CNOT gates into the central controlled gate $B$ for a three-qubit unitary is presented in Fig. \ref{fig:cnot-merging}.

\section{Circuit transpilation}
In this section, we present our method for transpiling the synthesized quantum circuit presented in Section \ref{sec:background} to a desired quantum hardware architecture. The goal of the transpilation process is to produce a quantum circuit such that all of the two-qubit gates are allowed on the target architecture, using as little extra gates as possible. In other words, a two-qubit gate on logical qubits $q_i$ and $q_j$ is allowed only if there exists a connection between the physical qubits $v_i$ and $v_j$ on the target hardware. From a practical perspective, it is equally important that the transpilation process can be performed quickly.

The optimized block-ZXZ decomposition described earlier is applied recursively until we are left with two-qubit unitaries, uniformly controlled $R_z$ gates and Hadamard gates. Since the decomposition yields $4^{n-2}$ two-qubit unitaries, it is imperative that the two physical qubits the unitaries are applied on are adjacent on the hardware, so that we do not need to use an exponential amount of extra gates to route them. Since the rest of the multi-qubit gates are uniformly controlled $R_z$ gates, we will need to be able to transpile them as efficiently as possible. The following three sections explain this procedure in detail: Section \ref{section:initial-mapping} describes the process of selecting the best initial physical qubits for the synthesis, Section \ref{section:gray-code-selection-and-swapping} explains how to construct the uniformly controlled $R_Z$ gates more efficiently by selecting a better Gray code and using SWAP gates, and finally, Section \ref{section:uc-gate-construction-and-cnot-merging} introduces a heuristic for reducing the two-qubit gate count further. An overview of the procedure can be found in Table \ref{tab:transpilation-overview}.

\subsection{Selecting the initial qubit mapping}
\label{section:initial-mapping}
We start by introducing the notation used in the procedure. We denote the underlying graph of the target hardware architecture as $G$. The physical qubits are denoted as $V$, and the coupling map (the connections between the physical qubits) as $E$. A uniformly controlled $R_z$ gate on target qubit $v$ is denoted by $UC_v$

Let $S=\{v_0, \cdots, v_{n-1}\} \subseteq V$ be the set of physical qubits we apply the $n$-qubit unitary on. In the first level of the recursion, the target qubit of the uniformly controlled $R_z$ gates is $v_0$. Since we recurse until we have only two-qubit unitaries left, the last target qubit of the $UC$ gates is $v_{n-3}$. As the $UC$ gate for $n$ qubits decomposes into $2^{n-1}$ CNOT gates between the target qubit and the control qubits \cite{multiplexer-decomp}, we must select and index the physical qubits such that the sum of all pairwise distances between the selected physical qubits is minimized. More formally, we wish to minimize the cost function:
\begin{align}
    \label{cdist}
    C_{dist} = \sum_{u \in S} \sum_{v \in S} d(u,v)
\end{align}

The minimum cost can be calculated exactly by trying every combination of $n$ nodes $v_0, \cdots, v_{n-1} \in V$. However, this approach scales exponentially as $ \binom{|V|}{n}=O(|V|^n)$ when $ \binom{|V|}{n}$ is considered as a function of $|V|$. As an example, with $n=6$ on a $|V|=156$-qubit architecture, $\binom{|V|}{n} > 1.8 \cdot 10^{10}$, which is not reflected very well for a time complexity of $O(n^2)$ if $\binom{|V|}{n}$ was considered as a function of $n$. Therefore, we only employ this strategy in the case of
\begin{align}
    \binom{|V|}{n} \leq 5 \cdot 10^4,
\end{align}
where the value $ 5 \cdot 10^4$ was selected heuristically. For larger graphs, we use a greedy strategy. Firstly, we notice that the number of $UC$ gates applied on a target qubit $v_i$ for $i \leq n-3$ produced by the recursion is given by:
\begin{align}
\#(UC_{v_i}) = \sum_{j=0}^i3\cdot 4^j
\label{eq:uc-gate-count}
\end{align}
Therefore, we must start the mapping procedure from the highest qubit index $n-3$, iteratively mapping the next qubit as close as possible to all the other previously selected qubits since the deeper we are in the recursion, the more $UC$ gates will be applied to that specific target qubit as in (\ref{eq:uc-gate-count}). We do this by first selecting an edge $\{u, v \} \in E$ such that the endpoints $u$ and $v$ have the best average closeness to the rest of the graph, given by the equation
\begin{align}
    \label{ccloseness}
    C_{closeness}(u, v) = \sum_{w \in V} \Big(d(u,w) + d(v,w) \Big) \hspace{3mm} \text{for } \{u,v\}\in E,
\end{align}
and setting $S=\{u,v\}$. This can be done using a breadth-first search in time $O(|V|^2)$. After that, we repeatedly add a node to $S$ that increases the total pairwise distance cost $C_{dist}$ the least. The greedy expansion can be performed in time $O(|V| \cdot n)$. Using these strategies, we can find a good set of physical qubits to apply the unitary gate decomposition on.

\subsection{Gray code selection and qubit swapping}
\label{section:gray-code-selection-and-swapping}
Selecting a good initial qubit mapping gives us the starting point for the transpilation process. After that, we can focus on how to apply the $UC$ gates on the selected physical qubits as cost efficiently as possible. First of all, we introduce the long range CNOT ladder which we employ for CNOT gates which need to be executed on nonadjacent physical qubits. Denote a CNOT gate with a control qubit $v_i$ and a target qubit $v_j$ by $\text{CNOT}^{v_i}_{v_j}$. Furthermore, denote the shortest path from physical qubit $v_i$ to $v_j$ in $G$ by $(u_0, \cdots, u_k)$, where $u_0 = v_i$ and $u_k = v_j$. Lastly, denote the distance between the CNOT control and target by $d(v_i, v_j) = k$. Then, for $k>1$, we define the long range CNOT ladder as
\begin{align}
\text{CNOT}^{v_i}_{v_j} &= \left( \stackrel{\longleftarrow }{\prod_{i=1}^{k-2}} \text{CNOT}^{u_i}_{u_{i+1}} \right) \cdot \left(\prod_{i=1}^{k-1} \text{CNOT}^{u_i}_{u_{i+1}} \right) \cdot \\
\nonumber
&\left( \stackrel{\longleftarrow }{\prod_{i=0}^{k-2}} \text{CNOT}^{u_i}_{u_{i+1}} \right) \cdot \left( \prod_{i=0}^{k-1} \text{CNOT}^{u_i}_{u_{i+1}} \right),
\end{align}
where $\stackrel{\longleftarrow }{\prod}$ denotes that the indexing runs in reverse order. Then, using this construction, we get the following CNOT gate count for the long range CNOT gate:
\begin{align}
    \text{Ladder CNOT count} = \begin{cases}
        4k-4 &\text{If } d(v_i, v_j)=k > 1 \\
        1 & \text{Otherwise}
    \end{cases}
    \label{eq:ladder-cost}
\end{align}

A visualization of the long range CNOT ladder for the example case $d(v_i, v_j)=3$ can be found in Fig. \ref{fig:cnot-ladder}.

\begin{figure}[!b]
    \centering
    \begin{quantikz}[row sep={0.6cm,between origins}, column sep = 0.1cm]
        \lstick{{$v_0$}}& \ctrl{1} & & & & \ctrl{1}& & & &\\
        \lstick{{$v_1$}}& \targ{} & \ctrl{1}& & \ctrl{1}& \targ{} & \ctrl{1}& & \ctrl{1}& \\
        \lstick{{$v_2$}}& & \targ{} &\ctrl{1}& \targ{} & & \targ{} & \ctrl{1}& \targ{} &\\
        \lstick{{$v_3$}}& & & \targ{} & & & &  \targ{} & &\\
    \end{quantikz}
    \caption{An example of a long range $\text{CNOT}^{v_0}_{v_3}$ gate. In this example, the shortest distance between the physical qubits $v_0$ and $v_3$ on the hardware is 3, and the shortest path is $(v_0, v_1, v_2, v_3)$.}
    \label{fig:cnot-ladder}
\end{figure}
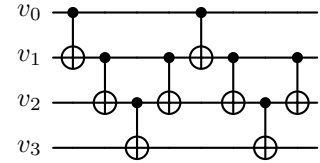

\begin{figure}[!b]
    \centering
    $a)$
    \hbox{
    \hspace{-5mm}\begin{quantikz}[row sep={0.6cm,between origins}, column sep = 0.1cm]
   \lstick{{$q_0$}}& & & & & & & & & \ctrl{3} & & & & & & & & \ctrl{3} &\\
   \lstick{{$q_1$}}& & & & & \ctrl{2} & & & & & & & & \ctrl{2} & & & & &\\
   \lstick{{$q_2$}}& & & \ctrl{1} & & & & \ctrl{1} & & & & \ctrl{1} & & & & \ctrl{1} & & &\\
   \lstick{{$q_3$}}& & \gate{R_z} & \targ{}& \gate{R_z} & \targ{} & \gate{R_z} & \targ{} & \gate{R_z} & \targ{} & \gate{R_z} & \targ{} & \gate{R_z} & \targ{} & \gate{R_z} & \targ{} & \gate{R_z} & \targ{} &
    \end{quantikz}} 
    $b)$
    \hbox{
    \hspace{-5mm}\begin{quantikz}[row sep={0.6cm,between origins}, column sep = 0.1cm]
   \lstick{{$q_0$}}& & & & & & & & & \ctrl{3} & & & & & & & & \ctrl{3} &\\
   \lstick{{$q_1$}}& & & \ctrl{2} & & & & \ctrl{2} & & & & \ctrl{2} & & & & \ctrl{2} & & &\\
   \lstick{{$q_2$}}& & & & & \ctrl{1} & & & & & & & & \ctrl{1} & & & & &\\
   \lstick{{$q_3$}}& & \gate{R_z} & \targ{}& \gate{R_z} & \targ{} & \gate{R_z} & \targ{} & \gate{R_z} & \targ{} & \gate{R_z} & \targ{} & \gate{R_z} & \targ{} & \gate{R_z} & \targ{} & \gate{R_z} & \targ{} &
    \end{quantikz}
    }
    \caption{Four-qubit uniformly controlled $R_z$ gates. a) The $UC$ gate constructed using a binary reflected Gray code as in \cite{multiplexer-decomp}. b) An equivalent construction for the $UC$ gate using a different Gray code. }
    \label{fig:different-gray-codes}
\end{figure}
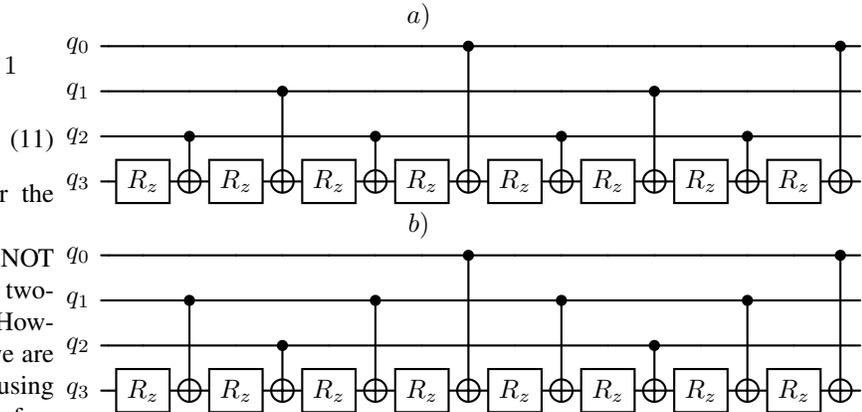
Using the initial qubit mapping and the long range CNOT ladders, we could construct the $UC$ gates such that all two-qubit operations are allowed within the target hardware. However, we can construct the $UC$ gates more efficiently if we are a bit more clever. In Fig. \ref{fig:different-gray-codes} a), the $UC$ gate is constructed using the binary reflected Gray code as in \cite{multiplexer-decomp}. However, we are free to choose \textit{any} cyclic Gray code which yields an equivalent $UC$ gate decomposition. An example of a different Gray code can be seen in Fig. \ref{fig:different-gray-codes} b). 

As mentioned in Section \ref{section:initial-mapping}, a $n$-qubit $UC$ gate consists of $2^{n-1}$ CNOT gates. For all valid cyclic Gray codes and for $n \geq 2$, the CNOT gates are grouped into $n-1$ frequencies $f_1, \dots f_{n-1}$. The CNOT gate count corresponding to a frequency $f_i$ is then given by

\begin{align}
    \label{eq:frequency}
    f_i = \begin{cases}
        2^{i-1} & \text{if } i>1 \\
        2 & \text{otherwise}.
    \end{cases}
\end{align}

As an example for the circuit in Fig. \ref{fig:different-gray-codes} a), qubits $q_0$ and $q_1$ have a CNOT frequency of 2, and qubit $q_2$ has a frequency of 4.

In order to minimize the extra CNOT gates induced by the usage of the long range CNOT ladders, we assign the highest frequency of CNOT gates $f_{n-1}$ in the cyclic Gray code to the physical qubits closest to the target, and the lowest frequency $f_1$ to the physical qubits furthest from the target. We do this by starting from the target qubit, and greedily add the nearest qubit to be the control qubit with the highest CNOT frequency. If we denote the amount of qubits used in the current recursion level by $m$, then this assignation can be done in time $O(m)$ with the greedy approach. However, in order to get the Gray code as an output of binary strings, we need to apply each of the CNOT gates iteratively. Since a $m$-qubit $UC$ gate decomposes into $2^{m-1}$ CNOT gates, this procedure can be executed in time $O(2^m)$.

In order to get an equivalent construction for the $UC$ gate using a different Gray code, we must also modify the transformation matrix $M$ described in \cite{multiplexer-decomp}, which is used to transform the multi-controlled $R_z$ gate angles to the $UC$ gate angles. The demultiplexing process from the optimized block-ZXZ decomposition gives us a block diagonal matrix
\[\begin{bmatrix}
    D & 0 \\
    0 &D^\dagger{}
\end{bmatrix} = \begin{bmatrix}
    R_z(\alpha_1) & \\
    & \ddots & \\
    & & R_z(\alpha_{2^{m-1}})^\dagger
\end{bmatrix} = UC_m,\]
which describes the $UC$ gate we wish to decompose. However, in order to get the correct $R_z$ gate angles for the decomposed $UC$ gate, we need to apply the following transformation:
\begin{align}
    M \begin{bmatrix}
        \alpha_1 \\
        \vdots \\
        \alpha_{2^{m-1}}
    \end{bmatrix} = \begin{bmatrix}
        \theta_1 \\
        \vdots \\
        \theta_{2^{m-1}}
    \end{bmatrix}, \hspace{5mm}M_{ij} = 2^{-(m-1)}(-1)^{b_{j-1}\cdot g_{i-1}},
\end{align}

where $b_m$ stands for the binary code for the current target qubit state, and $g_m$ stands for the Gray code being used. After acquiring our more optimal Gray code, we simply plug it in as $g_m$. More explicitly, the binary code is a sequence of binary strings corresponding to the integers $j\in[0,\dots,2^{m-1}-1]$, and the Gray code is some permutation of the binary code, also known as the \textit{phase polynomial} \cite{gray-synth}. As a running example for the circuits in Fig. \ref{fig:different-gray-codes}, the circuit in a) has the binary code $000, 001, \dots 110, 111$ and Gray code $000, 001, 011, 010, 110, 111, 101, 100$. The circuit in b) has the same binary code as in a), but has a different Gray code: $000, 010, 011, 001, 101, 111, 110, 100$.

This procedure gives us a more efficient way of constructing the $UC$ gates. However, we employ one more optimization strategy to get an even better decomposition: swapping. Denote the physical target qubit of the current recursion level by $v_0$, and the last physical target qubit at the second to last level of recursion (before we apply the two-qubit unitaries) by $v_m$. Denote the shortest path between these qubits by $(v_0, \cdots, v_m)$. Sometimes it is more efficient to swap the physical qubit $v_0$ closer to the last physical qubit $v_m$ using SWAP gates, since it might be closer to the rest of the physical qubits $v_1, \cdots, v_m \in S$ after the SWAP operation. Consequently, we need to use less of the expensive long range CNOT ladders. Thus, we iteratively swap the target qubit closer to the last target qubit, calculating the optimal Gray code greedily, and estimating the cost of the obtained $UC$ gate construction for each iteration using (\ref{eq:ladder-cost}) and (\ref{eq:frequency}). We then select the most efficient construction after adding the cost of the SWAP gates to the estimate.

\begin{figure}[!b]
    \centering
    a) \includegraphics[width=1\linewidth]{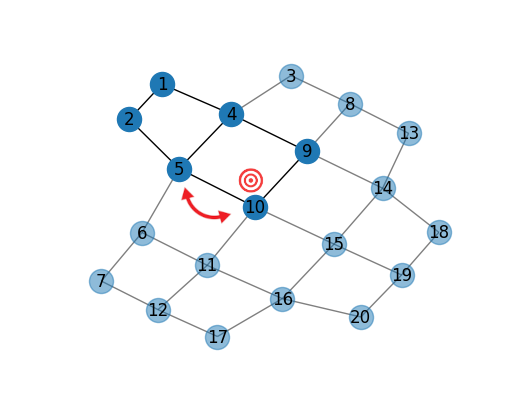}
    b) \includegraphics[width=1\linewidth]{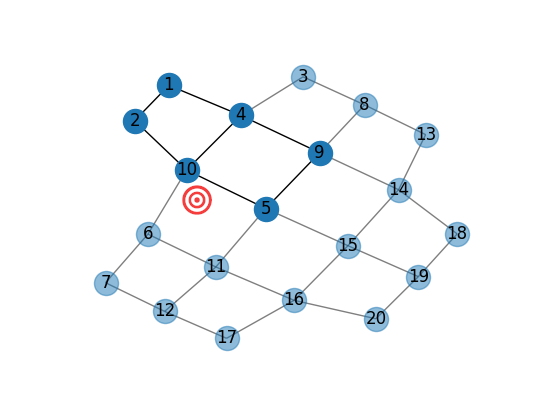}
    \caption{An example of the swapping procedure for a 6-qubit $UC$ gate on the IQM Garnet quantum hardware. a) The physical qubit 10 is selected as the target. Physical qubits 9, 5, 4, 2 and 1 are the control qubits of the $UC$ gate. b) Physical qubits 10 and 5 are swapped. This gives a more efficient construction for the $UC$ gate, as the target qubit 10 is now overall closer to the rest of the physical qubits.}
    \label{fig:swapping}
\end{figure}

As we swap at most $m$ times, calculate the optimal Gray code after each SWAP, and the optimal Gray code can be calculated and estimated in time $O(2^m)$, we get a time complexity of $O(m^2 \cdot 2^m)$ for the whole optimization process for one level of recursion. Even though the time complexity is exponential, we wish to point out that in practical terms usually $m < 14$, which comes from the fact that the unitary we are decomposing is usually smaller than $2^{14} \times 2^{14}$. An example of the swapping procedure can be seen in Fig. \ref{fig:swapping}.

\subsection{Uniformly controlled $R_z$ gate construction and CNOT merging}
\label{section:uc-gate-construction-and-cnot-merging}
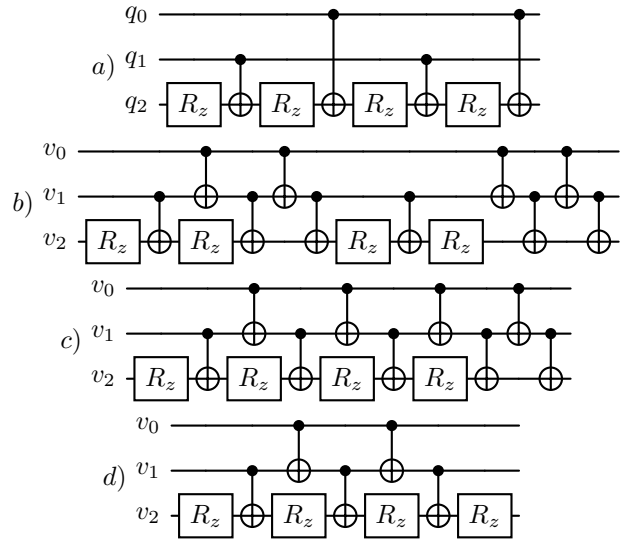
\begin{figure}[!b]
    \centering
    $a)$\begin{quantikz}[row sep={0.6cm,between origins}, column sep = 0.1cm]
        \lstick{{$q_0$}} & & & & \ctrl{2} & & & & \ctrl{2} &\\
        \lstick{{$q_1$}} & & \ctrl{1} & & & & \ctrl{1} & & &\\
        \lstick{{$q_2$}} & \gate{R_z} & \targ{} & \gate{R_z} & \targ{} & \gate{R_z} & \targ{} & \gate{R_z} & \targ{} &
    \end{quantikz} \\
    $b)$\begin{quantikz}[row sep={0.6cm,between origins}, column sep = 0.1cm]
        \lstick{{$v_0$}} & & & \ctrl{1} & & \ctrl{1} & & & & & \ctrl{1} & & \ctrl{1} & &\\
        \lstick{{$v_1$}} & & \ctrl{1} & \targ{} & \ctrl{1} & \targ{} & \ctrl{1} & & \ctrl{1} & & \targ{} & \ctrl{1} & \targ{} & \ctrl{1} &\\
        \lstick{{$v_2$}} & \gate{R_z} & \targ{} & \gate{R_z} & \targ{} & & \targ{} & \gate{R_z} & \targ{} & \gate{R_z} & & \targ{} & & \targ{}&
    \end{quantikz} \\
    $c)$\begin{quantikz}[row sep={0.6cm,between origins}, column sep = 0.1cm]
        \lstick{{$v_0$}} & & & \ctrl{1} & &  \ctrl{1} & & \ctrl{1} & & \ctrl{1} & &\\
        \lstick{{$v_1$}} & & \ctrl{1} & \targ{} & \ctrl{1} &  \targ{} & \ctrl{1} & \targ{} & \ctrl{1} & \targ{} & \ctrl{1} &\\
        \lstick{{$v_2$}} & \gate{R_z} & \targ{} & \gate{R_z} & \targ{} & \gate{R_z}  & \targ{} & \gate{R_z} & \targ{} & & \targ{} &
    \end{quantikz} \\
    $d)$\begin{quantikz}[row sep={0.6cm,between origins}, column sep = 0.1cm]
        \lstick{{$v_0$}} & & & \ctrl{1} & &  \ctrl{1} & & & \\
        \lstick{{$v_1$}} & & \ctrl{1} & \targ{} & \ctrl{1} &  \targ{} & \ctrl{1} & &\\
        \lstick{{$v_2$}} & \gate{R_z} & \targ{} & \gate{R_z} & \targ{} & \gate{R_z}  & \targ{} & \gate{R_z} & 
    \end{quantikz}
    \caption{Efficient construction of a three-qubit $UC$ gate using the long range CNOT ladders. a) The logical decomposition of the $UC$ gate, as in Fig. \ref{fig:cnot-merging}. b) The decomposition of the $UC$ gate on physical qubits $v_0$, $v_1$ and $v_2$ on a path graph using the binary reflected Gray code. c) A more efficient decomposition, using the heuristic of removing the nearest neighbor CNOT gate after a long range CNOT gate. d) The final decomposition after merging the rightmost CNOT gates into the neighboring unitary gate, as in Fig. \ref{fig:cnot-merging}.}
    \label{fig:efficient-ladder}
\end{figure}

After selecting the Gray code and the swap sequence for the $UC$ gate, we are almost ready to decompose it into elementary gates. However, we introduce the final heuristic which allows us to use the long range CNOT ladders more efficiently. In Fig. \ref{fig:efficient-ladder} a) we see the logical decomposition for a three-qubit $UC$ gate. Assume we wish to transpile this decomposition to three physical qubits $v_0$, $v_1$ and $v_2$ on a path graph $(v_0, v_1, v_2)$ using the binary reflected  Gray code. Fig. \ref{fig:efficient-ladder} b) shows how this is done naively using the long range CNOT ladders. As you notice, the "leg" of the long range CNOT ladder reaches the target qubit once before the whole CNOT gate is executed completely, entangling the qubits $v_0$, $v_1$ and $v_2$. We can use this to our advantage to save some CNOT gates, by applying the following heuristic:

\begin{itemize}
    \item[1.] When the "leg" of the CNOT ladder reaches the target qubit for the first time, apply a $R_z$ gate if the corresponding Gray code term has not been explored yet.
    \item[2.] When the long range CNOT gate has been executed completely, apply the next CNOT gate in queue. If the next CNOT gate is \textbf{not} a long range CNOT gate, and the resulting Gray code term \textit{has} been explored before, remove the local CNOT gate and continue.
\end{itemize}

As can be seen in Fig. \ref{fig:efficient-ladder} c), this heuristic saves some CNOT gates. We do not provide any further analysis for the correctness nor the CNOT gate reduction of the heuristic method. According to our experiments, the method works flawlessly most of the time. However, sometimes (with qubit counts of 12 and over), the method may fail to find some Gray code terms. In order to get a correct construction for the $UC$ gate, we employ GraySynth \cite{gray-synth} to find the missing terms with minimal number of extra CNOT gates. The extra gates are then appended to the end of the otherwise complete $UC$ gate circuit, ensuring correctness. Since the method uses simple local checks and one gate backtracking, it can be applied with the same time complexity as constructing the $UC$ gate in the first place. For a $m$-qubit $UC$ gate, this time complexity is again $O(2^{m})$.

After applying the heuristic method, we are left with the circuit in Fig. \ref{fig:efficient-ladder} c). Similarly to Fig. \ref{fig:cnot-merging}, we can merge the rightmost CNOT gates into the neighboring unitary gate. This gives us the final optimized $UC$ gate, which can be seen in Fig. \ref{fig:efficient-ladder} d). This concludes our optimization process for the $UC$ gates. For the current recursion level, this optimization process is dominated by the $O(m^2 \cdot 2^m)$ time complexity for the Gray code selection and swapping. Since we start the recursion from $n$ qubits, the final time complexity for the whole optimization process is $O(|V|^2+|V|\cdot n + n^2 \cdot2^n)$. As a remark, simply decomposing a $n$-qubit $UC$ gate logically takes $O(2^n)$ time. A Python code implementation for the full transpilation algorithm can be found in \url{https://github.com/QuantumHel/Architecture-aware-unitary-synthesis}.

\begin{table*}[!t]
\centering
\caption{Overview of the unitary synthesis and transpilation process}
\scalebox{1}{
\fbox{
\begin{tabular}{l}
{\bf Architecture-aware unitary synthesis} \\
\toprule
{\bf Input}
The unitary matrix $U \in U(2^n)$ to be synthesized and the coupling map $\{\{v_i,v_j\} \cdots \{v_k, v_l\}\}\in E$ \\
{\bf Initialization}\\
Select the best initial qubit mapping (Section \ref{section:initial-mapping}); \\
\bf {For $i$ in $[n, \cdots ,2]$:} \\
\ \ \ Calculate the optimal Gray code and set it as the current best (Section \ref{section:gray-code-selection-and-swapping}); \\
\ \ \ \bf {For $j$ in $[n, \cdots, 3]$:} \\
\ \ \ \ \ \ Swap the physical qubits corresponding to the logical qubits $q_j$ and $q_{j-1}$ (Section \ref{section:gray-code-selection-and-swapping}); \\
\ \ \ \ \ \ Calculate the optimal Gray code for the current layout (Section \ref{section:gray-code-selection-and-swapping});\\
\ \ \ \ \ \ If the current Gray code is more efficient than the current best, set the current Gray code as the best;\\
\ \ \  \bf {End} \\
\bf{End} \\
\bf{Recursion} \\
Decompose $U$ logically into the unitaries $W_C$, $V_A$, the central controlled unitary $B$, two $UC$ gates and two Hadamard gates (Section \ref{sec:background}); \\
Decompose the $UC$ gates with the best selected Gray code, applying the heuristic described in Section \ref{section:uc-gate-construction-and-cnot-merging} and respecting the coupling map; \\
Migrate the outer-most CNOT gates through the Hadamard gates, turning them into CZ gates (Fig. \ref{fig:cnot-merging}); \\
Merge the CZ gates into the central controlled unitary $B$ (Sections \ref{sec:background} and \ref{section:uc-gate-construction-and-cnot-merging}); \\
Decompose the central controlled gate $B$ into $W_B$ and $V_B$ (Section \ref{sec:background}); \\
Recurse for the unitaries $W_C, V_A, W_B$ and $V_B$ until only two-qubit unitaries are left (Section \ref{sec:background}); \\
\bf{Two-qubit unitary decomposition} \\
Starting from the last two-qubit unitary, decompose it up to a diagonal using two CNOT gates (Section \ref{sec:background}); \\
Migrate the diagonal through the controls of the $UC$ gates, and merge it with the next two-qubit unitary (Section \ref{sec:background}); \\
Continue this until only one two-qubit unitary is left. Decompose the last two-qubit unitary with three CNOT gates (Section \ref{sec:background}); \\
Decompose any one-qubit unitaries with the ZYZ-decomposition (Section \ref{sec:background});
\end{tabular}
}
}

\label{tab:transpilation-overview}
\end{table*}

\section{Results}
\label{sec:results}
To showcase the efficiency of our proposed transpilation method, we benchmarked it with random unitary matrices against three well known and optimized transpilers: Qiskit~\cite{qiskit2023}, TKET~\cite{sivarajah2021tket} and Pennylane~\cite{bergholm2018pennylane}. We recognize the existence of the well known and distinguished transpiler called BQSKit \cite{bqskit}, however, we decided not to include it in our benchmarks due to slow transpilation times. Furthermore, BQSKit transpiles the circuits up to arbitrary precision using numerical methods, as opposed to the other transpilers which create an exact equivalent circuit, making the comparisons ambiguous. Additionally, we point out that TKET and Pennylane do not offer unitary synthesis as part of their libraries. Thus, we create the logical circuit for the synthesized unitary in Qiskit, import it into TKET and Pennylane, and use their transpilers to transpile the circuit. All of the transpilers were run on the maximum optimization level.

We benchmarked the unitary synthesis transpilation on two different hardware architectures: the 20-qubit IQM Garnet \cite{iqm-garnet} and the 156-qubit IBM Marrakesh. The IQM Garnet machine has a square lattice topology, which is used in machines by other manufacturers as well, for example the Rigetti Ankaa-2 and the Google Sycamore \cite{google-sycamore}. The IBM Marrakesh has the heavy-hex topology, which is widely used in other IBM machines as well. Table \ref{tab:cnot-count-iqm} shows the CNOT count after the transpilation routine for the IQM Garnet with different transpilers, with the qubit count ranging from 3 to 11. Similarly, CNOT counts for the IBM Marrakesh can be seen in Table \ref{tab:cnot-count-ibm}. Furthermore, we present the transpilation time for each of the transpilers for the IQM Garnet in Table \ref{tab:runtime-iqm}, and for the IBM Marrakesh in Table \ref{tab:runtime-ibm}. We capped the transpilation time to 30 minutes, marking the entries with N/A which exceeded this time limit. All of the benchmarks were run on a Intel(R) Core(TM) i5-1135G7 @ 2.40GHz processor.
\begin{table*}[!th]
    \centering
    \caption{CNOT count for a random unitary synthesis on square lattice architecture (IQM Garnet) for different qubit counts}
    \begin{center}
    \label{tab:cnot-count-iqm}
        
    \begin{tabular}{|c|c|c|c|c|c|c|c|c|c|}
    \hline
     Transpiler / Num. qubits & 3 & 4 & 5 & 6 & 7 & 8 & 9 & 10 & 11 \\ 
     \hline
    Proposed method & \textbf{27} & \textbf{135} & \textbf{617} & \textbf{2599} & \textbf{10585} & \textbf{42801} & \textbf{172025} & \textbf{689873} & \textbf{2762999} \\
    \hline
    TKET & 31  & 147 & 651 & 2728 & 11237 & 45505 & N/A & N/A & N/A  \\
    \hline
    Qiskit & 29  & 183 & 903 & 3796 & 15606 & 61386 & 242271 & 957936 & N/A  \\
    \hline
    Pennylane & 40 & 221 & 963 & 4033 & 16517 & N/A & N/A & N/A & N/A  \\
    \hline
    \end{tabular}
    \end{center}

    \centering
    \caption{CNOT count for a random unitary synthesis on heavy-hex architecture (IBM Marrakesh) for different qubit counts}
    \begin{center}
    \label{tab:cnot-count-ibm}
        
    \begin{tabular}{|c|c|c|c|c|c|c|c|c|c|}
    \hline
     Transpiler / Num. qubits & 3 & 4 & 5 & 6 & 7 & 8 & 9 & 10 & 11 \\ 
     \hline
    Proposed method  & \textbf{27} & \textbf{145} & \textbf{681} & \textbf{2873} & \textbf{11761} & \textbf{49521} & \textbf{194697} & \textbf{757939} & \textbf{3020169} \\
    \hline
    TKET & 35  & 157 & 768 & 3684 & 15429 & 64508 & N/A & N/A & N/A  \\
    \hline
    Qiskit & 31  & 177 & 833 & 3658 & 15047 & 60748 & 244311 & 979867 & N/A \\
    \hline
    Pennylane & 37 & 221 & 1053 & 4573 & 19037 & N/A & N/A & N/A & N/A   \\
    \hline
    \end{tabular}
    \end{center}

    \centering
    \caption{Transpiler runtime for a random unitary synthesis on square lattice architecture (IQM Garnet) for different qubit counts on Intel(R) Core(TM) i5-1135G7 @ 2.40GHz processor}
    \begin{center}
    \label{tab:runtime-iqm}
        
    \begin{tabular}{|c|c|c|c|c|c|c|c|c|c|}
    \hline
     Transpiler / Num. qubits & 3 & 4 & 5 & 6 & 7 & 8 & 9 & 10 & 11\\ 
     \hline
    Proposed method & 0.02s & \textbf{0.04s} & \textbf{0.16s} & \textbf{0.61s} & \textbf{1.58s} & \textbf{6.43s} & \textbf{26.63s} & \textbf{115.48s} & \textbf{442.56s}\\
    \hline
    TKET & 3.4s  & 9.1s & 41.1s & 149.6s & 418.7s & 961.1s & N/A & N/A & N/A  \\
    \hline
    Qiskit & 5.54s  & 0.1s & 0.62s & 2.31s & 7.36s & 26.8s & 134.5s & 599.4s & N/A   \\
    \hline
    Pennylane & \textbf{0.0s} & 2.9s & 10.6s & 123.2s & 1100.3s & N/A & N/A & N/A & N/A   \\
    \hline
    \end{tabular}
    \end{center}

    \centering
    \caption{Transpiler runtime for a random unitary synthesis on heavy-hex architecture (IBM Marrakesh) for different qubit counts on Intel(R) Core(TM) i5-1135G7 @ 2.40GHz processor}
    \begin{center}
    \label{tab:runtime-ibm}
    \begin{tabular}{|c|c|c|c|c|c|c|c|c|c|}
    \hline
     Transpiler / Num. qubits & 3 & 4 & 5 & 6 & 7 & 8 & 9 & 10 & 11\\ 
     \hline
    Proposed method  & \textbf{0.02s} & \textbf{0.03s} & \textbf{0.12s} & \textbf{0.44s} & \textbf{1.73s} & \textbf{5.89s} & \textbf{28.4s} & \textbf{118.77s} & \textbf{499.77s}  \\
    \hline
    TKET & 18.8s  & 27.1s & 61.0s & 195.7s & 538.5s & 1224.8s & N/A & N/A & N/A  \\
    \hline
    Qiskit & 2.93s  & 3.03s & 3.23s &  5.1s & 12.23s & 31.21s & 132.6s & 617.2s & N/A  \\
    \hline
    Pennylane & 9.1s & 4.3s & 14.4s & 104.3s & 1374.3s & N/A & N/A & N/A & N/A \\
    \hline
    \end{tabular}
    \end{center}
    
\end{table*}
\begin{table*}[]
    \centering
    \caption{CNOT count \textbf{increase} of proposed method compared to the theoretical lower bound on different architectures}
    \begin{center}
    \label{tab:ratio-theoretical}
    \begin{tabular}{|c|c|c|c|c|c|c|c|c|c|}
    \hline
     Architecture / Num. qubits & 3 & 4 & 5 & 6 & 7 & 8 & 9 & 10 & 11\\ 
     \hline
    IQM Garnet & 42.1\% & 42.1\% & 45.9\% & 45.9\% & 44.6\% & 44.3\% & 44.1\% & 44.0\% & 44.0\%\\
    \hline
    IBM Marrakesh & 42.1\% & 52.6\% & 61.0\% & 61.1\% & 60.7\% & 67.0\% & 63.1\% & 58.2\% & 57.4\% \\
    \hline
    \end{tabular}
    \end{center}
    \centering
    \caption{CNOT count \textbf{decrease} of proposed method compared to Qiskit, TKET and Pennylane on square lattice architecture (IQM Garnet) for different qubit counts}
    \begin{center}
    \label{tab:ratio-iqm}
    \begin{tabular}{|c|c|c|c|c|c|c|c|c|}
    \hline
     Transpiler / Num. qubits & 3 & 4 & 5 & 6 & 7 & 8 & 9 & 10  \\ 
     \hline
    TKET & 12.9\%  & 8.2\% & 5.2\% & 4.7\% & 5.8\% & 5.9\% & N/A & N/A   \\
    \hline
    Qiskit & 6.9\%  & 26.2\% & 31.7\% & 31.5\% & 32.2\% & 30.3\% & 29.0\% & 28.0\% \\
    \hline
    Pennylane & 32.5\% & 38.9\% & 35.9\% & 35.6\% & 35.9\% & N/A & N/A & N/A   \\
    \hline
    \end{tabular}
    \end{center}
    \centering
    \caption{CNOT count \textbf{decrease} of proposed method compared to Qiskit, TKET and Pennylane on heavy-hex architecture (IBM Marrakesh) for different qubit counts}
    \begin{center}
    \label{tab:ratio-ibm}
    \begin{tabular}{|c|c|c|c|c|c|c|c|c|}
    \hline
     Transpiler / Num. qubits & 3 & 4 & 5 & 6 & 7 & 8 & 9 & 10  \\ 
     \hline
    TKET & 22.9\%  & 7.60\% & 11.3\% & 22.0\% & 23.8\% & 23.2\% & N/A & N/A   \\
    \hline
    Qiskit & 12.9\%  & 18.1\% & 18.2\% & 21.5\% & 21.8\% & 18.5\% & 20.3\% & 22.6\% \\
    \hline
    Pennylane & 27.0\% & 34.4\% & 35.3\% & 37.2\% & 38.2\% & N/A & N/A & N/A   \\
    \hline
    \end{tabular}
    \end{center}
\end{table*}
\section{Discussion}
From Tables \ref{tab:cnot-count-iqm}-\ref{tab:runtime-ibm} we can see that our proposed transpilation method beat every other transpiler in the CNOT count \textit{and} runtime. Based on the results, the TKET transpiler achieves the next best CNOT count for qubit counts of over 3 on the IQM Garnet. Conversely, the Qiskit transpiler achieved better CNOT counts on the IBM Marrakesh architecture for qubit counts larger than 5. On the other hand, the Qiskit transpiler is significantly faster compared to TKET and Pennylane for both architectures. Neither TKET nor Pennylane could transpile circuits with over eight qubits in under 30 minutes. In order to run quantum circuits in practice, it is important that the transpilation process is done fast, so that one can actually gain some benefits from running the circuit on a quantum computer in the first place. However, speed alone is not enough to achieve practically viable quantum circuits; compared to Qiskit, our method achieved up to 32\% decrease in CNOT count when transpiling to the IQM Garnet.

On the IQM Garnet, our proposed method achieved an average CNOT count reduction of 7\% compared to TKET, 27\% compared to Qiskit and 36\% compared to Pennylane. On the IBM Marrakesh, the average reduction was 18\% when compared to TKET, 19\% compared to Qiskit and 34\% compared to Pennylane. In terms of transpiler runtime, we achieved a 260x average speedup compared to TKET, 39x speedup compared to Qiskit and 220x speedup compared to Pennylane on the IQM Garnet. On the IBM Marrakesh, the average speedup was 553x compared to TKET, 39x when compared to Qiskit and 350x when compared to Pennylane. Thus, our method is superior for both CNOT count and runtime. As mentioned in Section \ref{sec:results}, we did consider BQSKit for benchmarking as well. Our experiments showed that their transpiler was capable of outperforming our method in terms of CNOT count. However, they use numerical methods for the transpilation, and their transpilation time often exceeded the 30-minute time limit, therefore not ensuring an equitable comparison.

According to our experiments, applying only the initial qubit mapping (Section \ref{section:initial-mapping}), and the Gray code selection and qubit swapping (Section \ref{section:gray-code-selection-and-swapping}) gives us similar CNOT gate counts as the other transpilers we benchmarked against. Therefore, the heuristic $UC$ gate construction and CNOT gate merging techniques described in Section \ref{section:uc-gate-construction-and-cnot-merging} are crucial in order to push the gate count reduction over the capabilities of the other transpilers. We mentioned in Section \ref{section:uc-gate-construction-and-cnot-merging} that for qubit counts of 12 and over, the $UC$ gate construction procedure might require a fallback to the GraySynth in order to fill in some missing terms of the $UC$ gate. However, this is not a problem since at 11 qubits, there are already millions of CNOT gates, which is way beyond the NISQ regime. Therefore, the extra gates produced by the GraySynth fallback are negligible compared to the overall gate count. As an example, for the IBM Marrakesh at 12 qubits, the heuristic misses 21 terms in total. Assuming the absolute worst case scenario, the GraySynth will add $21 \cdot (n-1) \cdot (4\cdot(n-1) - 4) = 9240$ CNOT gates, where $(n-1)$ denotes the maximum amount of CNOT gates required to change any $(n-1)$-bit binary string to any other binary string, and $(4 \cdot (n-1)-4)$ denotes the maximum distance between physical qubits in a $n$-qubit layout. As the 12-qubit unitary transpiles into 12802461 CNOT gates on the IBM Marrakesh, the CNOT gates added by the GraySynth only account for a fraction of $\frac{9240}{12802461} \approx 0.072 \%$ of the total CNOT count.

As stated in the introduction, the theoretical lower bound  for the number of CNOT gates required to implement a general $n$-qubit unitary is $\frac{1}{4}(4^n-3n-1)$. Table \ref{tab:ratio-theoretical} shows the CNOT gate ratio of our method compared to the theoretical bound for the IQM and IBM architectures. Calculating the mean gives us an average \textbf{increase} of 44.1\% on the IQM Garnet with a standard deviation of 1.27\%, and an average increase of 58.1\% with a standard deviation of 6.79\% for the IBM Marrakesh, which indicates that the routing cost is more or less constant for both architectures. We also notice that the average ratio for the IBM architecture is slightly higher than on the IQM architecture. This is explained by the fact that the heavy-hex topology has lower average connectivity than the square lattice, so longer routing paths are expected. Additionally, Tables \ref{tab:ratio-iqm} and \ref{tab:ratio-ibm} show the \textbf{decrease} in CNOT count when comparing our method with the other transpilers on different architectures. Calculating the means, we get 7.12\%, 27.0\% and 35.8\% with standard deviations 2.8\%, 7.8\% and 2.0\% for TKET, Qiskit, respectively and Pennylane on the IQM Garnet. For the IBM Marrakesh, we get the means 18.5\%, 19.2\% and 34.4\% with standard deviations 6.5\%, 2.9\% and 3.9\% for TKET, Qiskit and Pennylane, respectively. The stabilization of the CNOT count decrease further reinforces the finding that the overhead ratio is roughly constant, and persists as circuits scale.

With current two-qubit gate fidelities (99.29\% for the IQM Garnet at the time of writing), a circuit with $\approx$ 2,600 CNOT gates would essentially have zero fidelity, which is practically infeasible. However, recent improvements have pushed the fidelity up to 99.99\% for trapped-ion machines \cite{two-qubit-gate-fidelity}, giving a fidelity of $\approx$ 77\% for a 2,600 CNOT circuit. As the industry is pushing towards hitting the "four-nines" $(>99.99\%)$ consistently, these longer circuits may become practical reality in the upcoming years.

While our benchmarks span 3 to 11 qubits, it is worth being explicit about why this range constitutes the practically relevant regime for exact general unitary synthesis, rather than a limitation to be lifted by further experiments. The CNOT count of the block-ZXZ decomposition scales as $\Theta(4^n)$, so each additional qubit roughly quadruples the number of two-qubit gates. At 11 qubits the synthesized circuit already contains millions of CNOT gates, and by 12 qubits even our fast transpilation routine becomes dominated by the $O(2^n)$ cost of simply writing out the decomposition, well before execution on hardware with sub-99.99\% two-qubit fidelities becomes a consideration. The input itself becomes the bottleneck before transpilation is even relevant: representing a general $n$-qubit unitary requires a dense $2^n \times 2^n$ matrix, which is memory-prohibitive on classical hardware well before circuit depth is. Exact general unitary synthesis is therefore best understood as a tool for compiling small, dense unitary blocks that recur as subroutines inside much larger circuits (oracles, block-encodings, local error-correction encoding circuits), rather than as a method for compiling an entire algorithm's unitary at scale. Extending the empirical results past 11 qubits would mainly reconfirm this asymptotic behavior, since the scaling is dictated by the theoretical bound in (\ref{block-zxz-bound}) rather than by our transpilation strategy. The range we report was chosen to characterize the crossover into this impractical regime rather than to push the method's ceiling.

A related point concerns how this contribution fits within the broader compilation landscape. Because our method targets exact synthesis of dense, unstructured unitaries, it is necessarily specialized relative to approximate synthesis approaches such as BQSKit \cite{bqskit}, or algorithm-level compilation strategies that exploit problem-specific structure, for example QSP/QSVT constructions \cite{gilyen2019quantum} or oracle-specific synthesis, to avoid ever materializing a dense unitary at all. For large-scale near-term applications, such structure-aware methods will typically remain the more practical choice, since they sidestep the $\Theta(4^n)$ scaling that exact synthesis is subject to. We view our method as complementary to these approaches rather than competing with them: in workflows where an algorithm decomposes into a sequence of small, dense unitary blocks, for instance, local encoding unitaries in state preparation \cite{plesch-state-prep}, small block-encodings inside a larger QSVT circuit or the diagonalizing unitaries in Trotter-based time evolution \cite{wierichs2025unitarysynthesisoptimalbrick}, an algorithm-level compiler could dispatch each block to our method for exact, hardware-aware synthesis while relying on approximate or structural techniques for the surrounding circuit. This division of labor may be particularly relevant in the early fault-tolerant era, where logical qubit counts are small due to the overhead of QEC encodings, and where the cost of exact synthesis is more easily justified than in NISQ settings, since it avoids the additional logical error that approximate synthesis would otherwise introduce on already scarce logical qubits. Exploring this hybrid regime, where our transpilation method would be embedded as a subroutine within a larger, approximation-tolerant compiler, is a natural direction for future work.

Finally, our current formulation assumes a static coupling map $E$ with uniform edge costs, using the unweighted graph distance $d(u,v)$ throughout the cost functions in (\ref{cdist}) and (\ref{ccloseness}), as well as in the Gray code selection and swapping heuristics of Section \ref{section:gray-code-selection-and-swapping}. Real superconducting devices exhibit calibration drift, however, two-qubit gate and readout error rates fluctuate between calibration cycles, so the best physical qubits or edges to route through can change even when the topology itself does not. Because our qubit mapping and Gray code selection are recomputed from scratch for each synthesis call rather than fixed against a device snapshot, accommodating calibration drift would not require restructuring the algorithm: replacing the graph distance $d(u,v)$ with a fidelity-weighted distance, derived from the backend's current calibration data, in $C_{dist}$ and $C_{closeness}$, and in the CNOT-frequency assignment of Section \ref{section:gray-code-selection-and-swapping}, would let the same greedy mapping and Gray-code machinery preferentially route through higher-fidelity qubits and connections. Since these cost computations are already the dominant but still polynomial-time components of the pipeline (Section \ref{section:initial-mapping}), incorporating live calibration data would not change the method's asymptotic complexity. We have not evaluated this fidelity-aware variant empirically. Doing so, together with characterizing how often recomputation would be needed to track realistic drift timescales, is left for future work.

\section{Conclusions}

In this article, we have presented a novel architecture-aware transpilation method for exact general unitary gate synthesis that supports arbitrary quantum hardware topologies. The method is tightly integrated with the optimized block-ZXZ decomposition, exploiting its recursive structure to make hardware-conscious decisions at every level of the recursion. Our approach combines three complementary techniques: a greedy qubit mapping strategy that minimizes pairwise distances on the target coupling graph, an adaptive Gray code selection with qubit swapping that optimizes the construction of uniformly controlled $R_z$ gates for the physical topology, and a heuristic for reducing CNOT gates by leveraging the intermediate entangling steps of long range CNOT ladders.

We benchmarked our method against three well-established and optimized transpilers---TKET, Qiskit, and Pennylane---on two representative superconducting architectures: the 20-qubit IQM Garnet (square lattice) and the 156-qubit IBM Marrakesh (heavy-hex). Our method consistently achieved the lowest CNOT counts across all tested qubit counts and both architectures, with average reductions of up to 27\% on the IQM Garnet and up to 19\% on the IBM Marrakesh compared to Qiskit, and up to 36\% and 34\%, respectively, compared to Pennylane. Equally important, our transpilation runtime was orders of magnitude faster than the competing methods, with speedups of up to 553x compared to TKET and up to 39x compared to Qiskit. Notably, our method was the only transpiler capable of handling circuits beyond 10 qubits within a 30-minute time limit on both architectures, successfully transpiling circuits of up to 11 qubits.

Both the CNOT count and the transpilation speed are critical for the practical utility of quantum circuits. Excessive two-qubit gate counts introduce prohibitive noise on current hardware, while slow transpilation times can negate the computational advantage that quantum computing aims to provide. By addressing both of these concerns simultaneously, our method represents a meaningful step toward making exact unitary synthesis practically viable on near-term quantum devices.

We note that while several other approaches to quantum circuit optimization and topology-aware synthesis exist in the literature~\cite{circuit-partitioning, peephole, pcoast, topology}, direct comparison is difficult because they either do not support architecture-aware synthesis~\cite{pcoast}, rely on numerical approximation~\cite{topology, bqskit}, or lack publicly available implementations~\cite{circuit-partitioning, peephole}. Our implementation is publicly available at \url{https://github.com/QuantumHel/Architecture-aware-unitary-synthesis}.

Future work will focus on integrating our transpilation method into broader quantum algorithmic workflows. In particular, unitary synthesis is a core subroutine in quantum state preparation~\cite{plesch-state-prep}, and our method could be directly applied to reduce the hardware overhead of state preparation circuits. Other promising directions include extending the method to accommodate native gate sets beyond CNOT-based decompositions, exploring the interplay between our transpilation strategy and quantum error mitigation techniques, and investigating how the method scales on emerging hardware topologies with higher connectivity.

\newpage
\bibliography{ref}
\label{pages:refs}

\end{document}

%% file: styles.tikzstyles
\tikzstyle{X Spider}=[fill={rgb,255: red,232; green,165; blue,165}, draw=black, shape=circle, tikzit fill={rgb,255: red,232; green,165; blue,165}, tikzit draw=black]
\tikzstyle{Z Spider}=[fill={rgb,255: red,216; green,248; blue,216}, draw=black, shape=circle, tikzit fill={rgb,255: red,216; green,248; blue,216}, tikzit draw=black]
\tikzstyle{Hadamard}=[fill=yellow, draw=black, shape=rectangle, tikzit draw=black, tikzit fill=yellow]
